\documentclass{ws-procs9x6}

\begin{document}

\title{Happy Island\footnote{Invited talk presented at Symposium on Nuclear Physics in GOA, India, Nov. 28-Dec. 2, 2010}}

\author{Larry McLerran}

\address{Brookhaven National Laboratory and Riken Brookhaven Center,\\
Upton, NY, 11973-5000, USA\\}

\begin{abstract}
I discuss the phase diagram for QCD in the baryon chemical potential
and temperature plane.  I argue that there is a new phase of matter different from the deconfined Quark Gluon Plasma: Quarkyonic Matter.  Quarkyonic Matter is confined and exists at densities parametrically large compared to the QCD scale, when the number of quark colors, $N_c$ is large.  I motivate the possibility that Quarkyonic Matter is in an inhomogeneous phase, and is surrounded by lines of phase transitions, making a Happy Island in the $\mu_B$-T plane.  I conjecture about the geography of Happy Island. \end{abstract}

\section{Quarkyonic Matter}

In a typical phase diagram of QCD, one will see two phases: Confined Matter, and the Deconfined Quark Gluon Plasma.\cite{Cabibbo:1975ig}  In some more detailed cartoons, one will also see Color Superconductivity and the Liquid Gas Nuclear matter phase transition.\cite{Alford:1997zt},\cite{Rapp:1997zu} The latter two phases make small changes to the bulk properties of nuclear matter, and I shall not discuss them in detail below.  The transition between confined matter and deconfined matter is however a big deal, because the number of degrees of freedom changes rapidly.  At low temperature and density, the important degrees of freedom are pions, $N_{dof} =3$.  In deconfined matter,
there are 16 gluon and around $20-30$ quark degrees of freedom.  The energy density scaled by the temperature to the fourth power, $\epsilon/T^4$ is proportional to
the number of degrees of freedom and therefore makes a large change.  This change takes place in a relatively narrow range of temperature, $\Delta T \sim 30-40~MeV$.  

The arguments for such a a confinement-deconfinement transition are strong at finite temperature and zero baryon chemical potential.  There is abundant lattice Monte Carlo data that support such a picture.\cite{Boyd:1996bx}  On the other hand at finite baryon number chemical potential, little is known.  It is assumed by many that in the phase diagram there remains a a rapid transition between a confined and a deconfined world.

Rob Pisarski and I have recently provided arguments that the phase structure at finite density is quite different from this simple picture. \cite{McLerran:2007qj}, \cite{Hidaka:2008yy} To understand the physics issues consider a world where the number of quark colors is large.  In such a world, the baryon mass is $M_{baryon} \sim N_c \Lambda_{QCD}$, where $\Lambda_{QCD} \sim 200~MeV$ is the dimensional scale of QCD.  At low temperature and density, there are no baryons in the system since $e^{(\mu_B-M_N)/T} \sim e^{-N_c}$, for $\mu_B < M_N$.  Also, the low temperature world is always confined.  This is because in the limit of a large number of colors, quark loops are not important and do not contribute to the potential measured as a function of separation for two heavy quarks used as probes.  Since quark loops do not affect the confining potential, the deconfinement temperature is independent of $\mu_B$.
\begin{figure}[t]
 \center{\vskip 0in \hskip 0in\includegraphics[width=8cm]{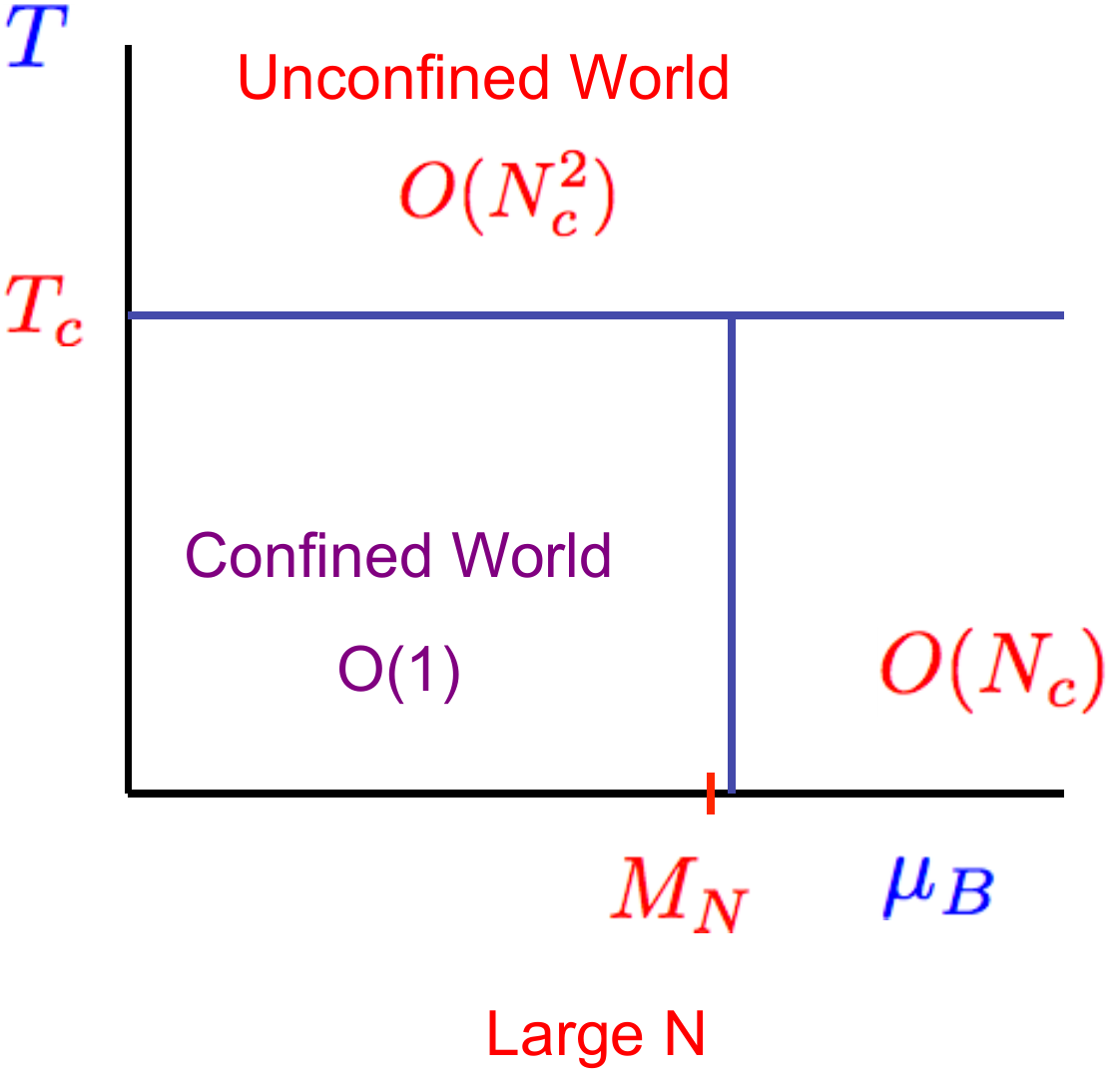} \vskip 0in }
 \caption{\label{fig_scales} The phase diagram of QCD in the large $N_c$ limit.  }
 \label{pd}
 \end{figure}

The resulting phase diagram is shown in Fig. \ref{pd} . Note that the threshold for high density matter is shifted from the nucleon mass due to interactions of nucleons.  The number of degrees of freedom in the unconfined world is of order $N_c^2$ since gluons have this number of degrees of freedom.  The confined world has $O(1)$ degrees of freedom.  Quarkyonic matter has only $N_c$ degrees of freedom, because gluons are permanently confined into glueballs.  

Quark-holes pairs and gluons are bound into hadrons in Quarkyonic matter.  Also, baryonic excitations near the Fermi surface should be bound into baryons.  However deep in the Fermi sea, the interactions of quarks are energetic and the confining part of the interaction should not play a significant role.  We may think
of the system approximately as a Fermi gas of free quarks, with all thermal and Fermi surface excitations permanently confined.  The name Quarkyonic is given because this matter has both the properties of free quarks, and confined baryons.

Generation of mass is associated with the breaking of chiral symmetry.  In the next section we shall argue how chiral symmetry breaking and restoration appears in the presence of Quarkyonic matter.  

\begin{figure}[t]
 \center{\vskip 0in \hskip 0in\includegraphics[width=10cm]{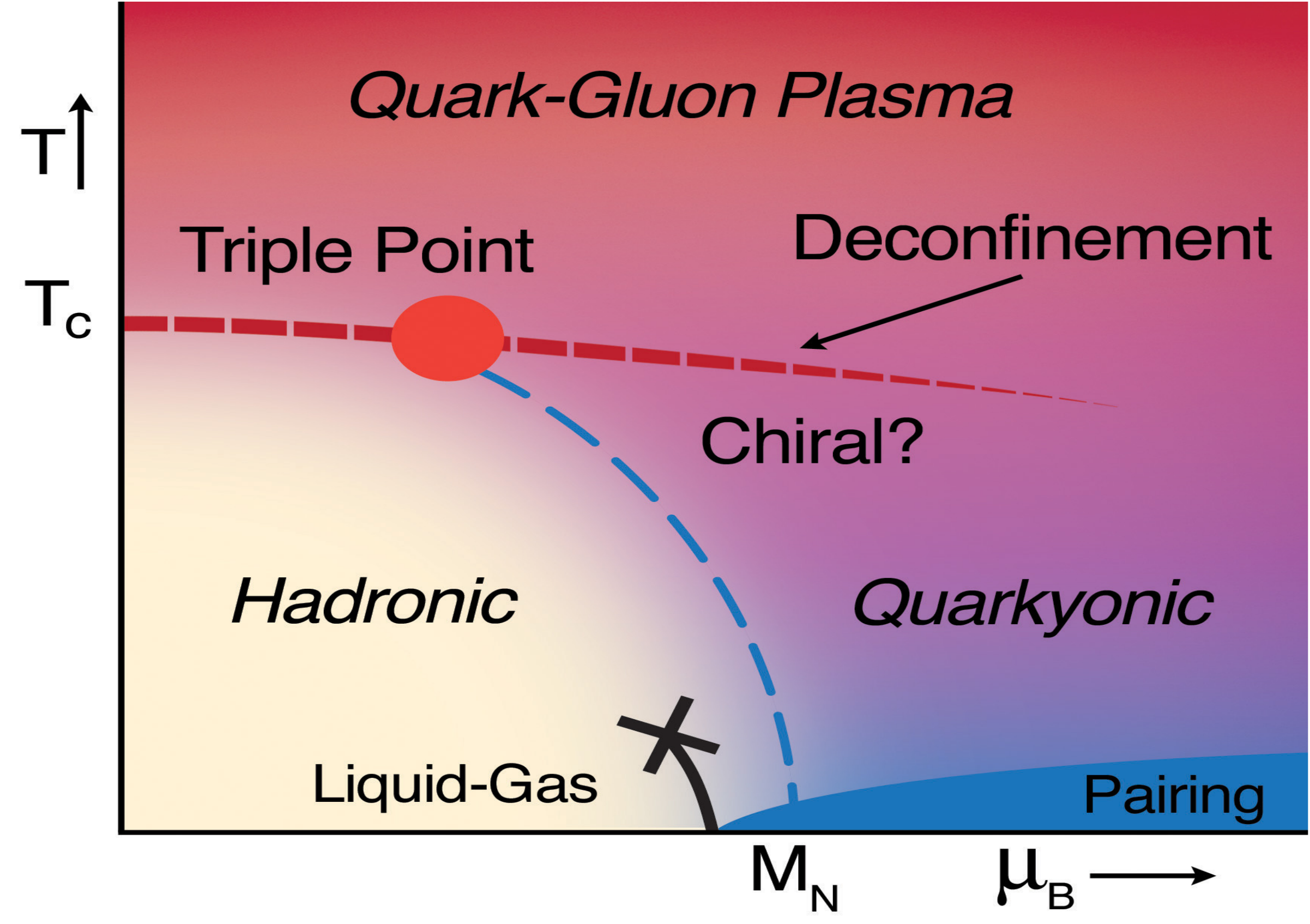} \vskip 0in }
 \caption{\label{fig_scales}  A somewhat realistic phase diagram of QCD  }
 \label{realpd}
 \end{figure}

A somewhat realistic plot of what we expect for $N_c = 3$ is shown in Fig. \ref{realpd}.
 
 Because of the three phases, there should be an approximate triple point in the QCD phase diagram.  this is the analog of the liquid-gas-solid triple point of water.  The triple point may or may not be a critical end point, depending upon dynamical details.  The triple point should correspond to rapid changes in the properties of the system as approached in any direction in the $\mu_B-T$ plane.
 
 \section{Chiral Symmetry Breaking and the Emergence of Happy Island}
 
 Early mean field studies of chiral symmetry breaking suggested that chiral symmetry should be restored in Quarkyonic Matter.\cite{Glozman:2008ja}\cite{McLerran:2008ua}  These studies assumed a spatially homogenous chiral condensate.  At finite density, one should in fact expect chiral condensates that break translational invariance.  
 
 Chiral condensation is generated by pairing between quark and hole pairs.  At zero density and temperature, the hole is in the negative energy Dirac sea and corresponds to an anti-quark.  At finite density, the same pairing can occur, with little cost in energy, if the quark has an energy a little above the Fermi surface, and the hole has an energy a little below the Fermi surface.  The energy loss, for massless quarks, in extracting such a pair is the same as the vacuum situation at low temperature and density.
 
 Such a quark-quark hole pair has an energy close to $2\mu_B$ since they both come from near the Fermi surface.  Pair formation into a meson state can take place if the pair has small relative momentum but total momentum near $2\mu_B$.  The condensate takes place into modes of high net momentum.  The pair has a finite DeBroglie wavelength, and therefore its condensation will involve the breaking of translational invariance.  This is the mechanism of charge density waves familiar in condensed matter physics.
 
 It can be argued that the dynamics of pair formation is essentially 1+1 dimensional, and that the dynamics should reduce to the 1+1 dimensional 't Hooft model.\cite{Kojo:2009ha}  There is a $2N_F$ Goldstone symmetry, where the extra factor of 2 arises from a degeneracy under spin.  The chiral condensation occurs through a rotation in space of $\overline \psi \psi$ with $\overline \psi \sigma^{0i} \psi$ where $i$ is in the direction of condensation.  Such a condensation macroscopically breaks parity.   This condensation is referred to as the chiral spiral.
 
 The formation of the condensation in the end is probably not 1 dimensional.\cite{Kojo:2010fe}  The Fermi surface breaks into patches with differing spatial orientations of the Fermi surface. The number of patches increases with density.
 
 It might also be possible that there are other types of translationally non-invariant condensates that form.
 this is suggested from a comprehensive analysis of mean field actions.\cite{Buballa:2009ct}
 
 The breaking of translation invariance signals a real phase transition, since an order parameter acquires an expectation value.  This means that there is probably some isolated island in the $\mu_B -T$ plane.
 
 This island is called Happy Island.  On the left hand side os Happy island, there is a sharp change in the number of degrees of freedom.  We would expect this part of Happy Island to have cliffs.  On the right hand side of Happy Island, there should be a relatively weak confinement-deconfinement phase transition, corresponding to beaches.  As one walks across Happy Island climbing from the beaches to the cliffs, one might have stair cases corresponding to the phase transitions associated with the patch structure on the Fermi surface. 
 
 Of course, history has shown that the discovery of new land masses is often done for reasons that turn out later to be not quite correct.  In this spirit one should expect surprises as we explore more.
 
 \begin{figure}[t]
 \center{\vskip 0in \hskip 0in\includegraphics[width=10cm]{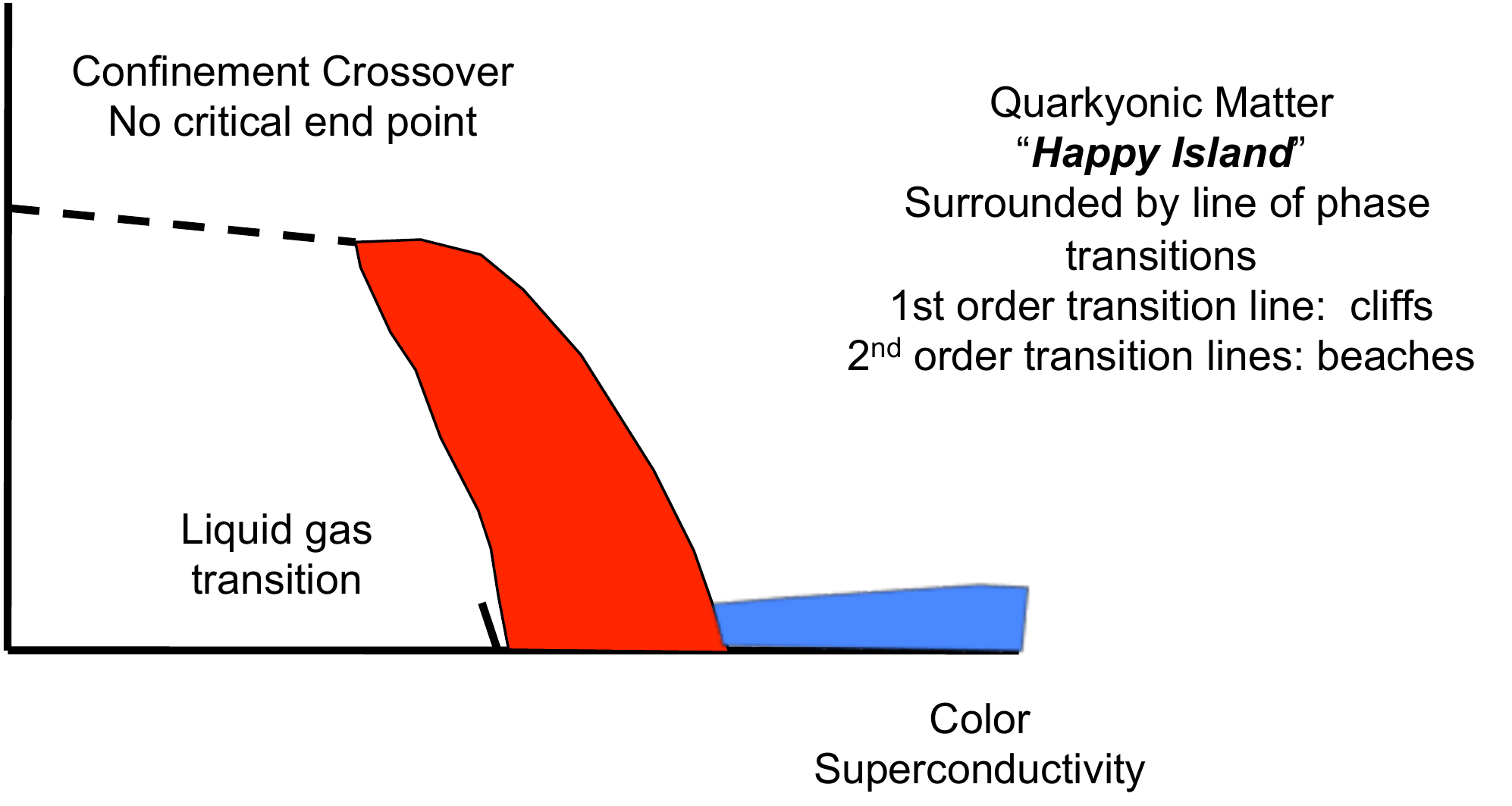} \vskip 0in }
 \caption{\label{fig_scales} Happy Island in the $\mu_B-T$ plane.  }
 \label{pd}
 \end{figure}

\section{Experimental Hints for Happy Island}

The existence of Quarkyonic Matter leads to the prediction of a triple point in the $\mu_B-T$ plane.
Also freeze out curves should parallel the boundary of Quarkyonic matter at low T and high $\mu_B$, and parallel the deconfinement transition at low $\mu_B$ and high T.  This is because along these transition lines,
the energy density jumps by roughly an order of magnitude in a narrow rand of either $T$ or $\mu_B$.  Thus in expansion, the matter may freeze out at $\mu_B$ and $T$ values characteristic of such a transition.

\begin{figure}[t]
 \center{\vskip 0in \hskip 0in\includegraphics[width=10cm]{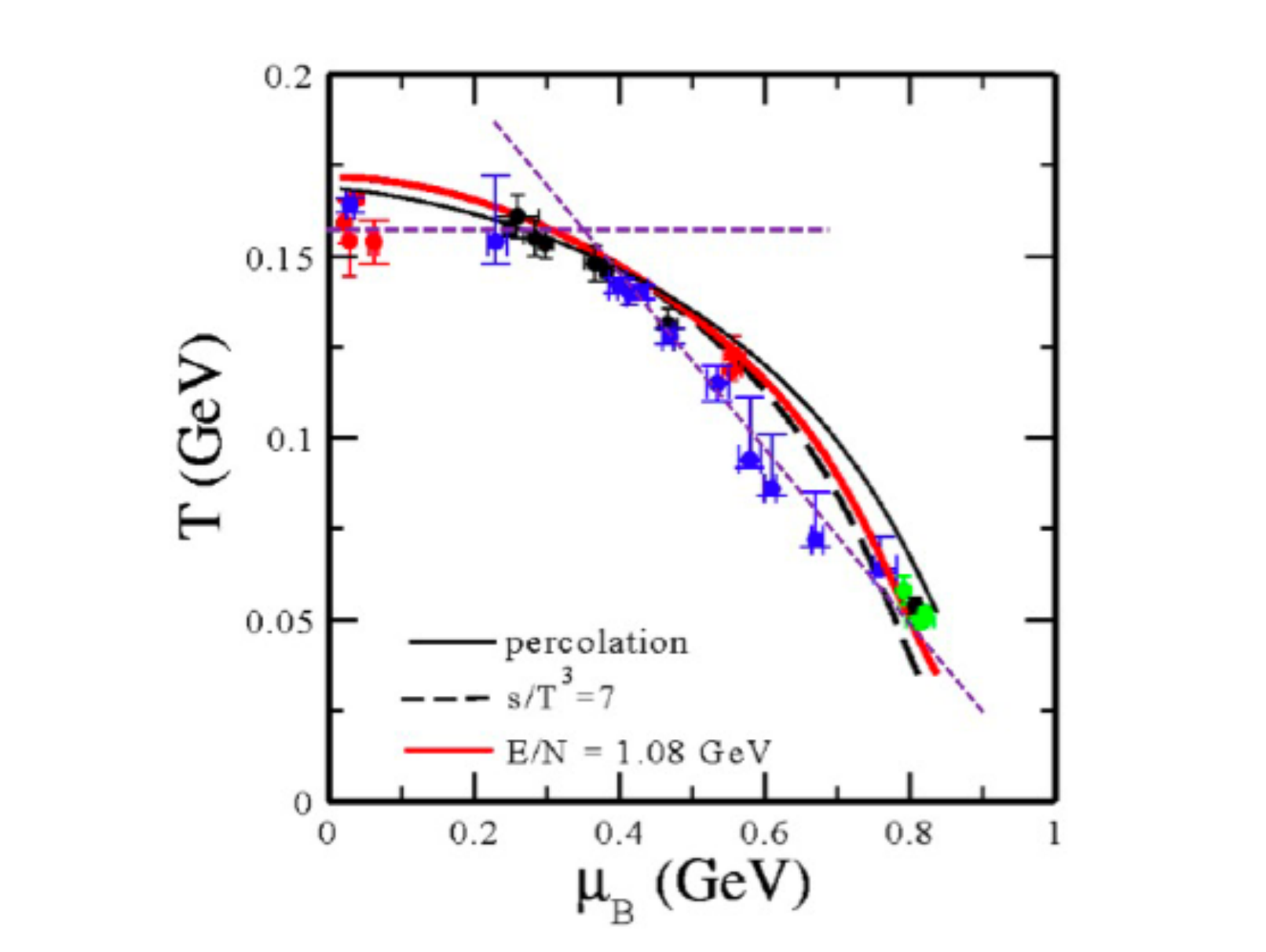} \vskip 0in }
 \caption{\label{fig_scales} The freeze out curve for $\mu_B$ and $T$ compared to a simple model.  }
 \label{freezeout}
 \end{figure}

\begin{figure}[t]
 \center{\vskip 0in \hskip 0in\includegraphics[width=12cm]{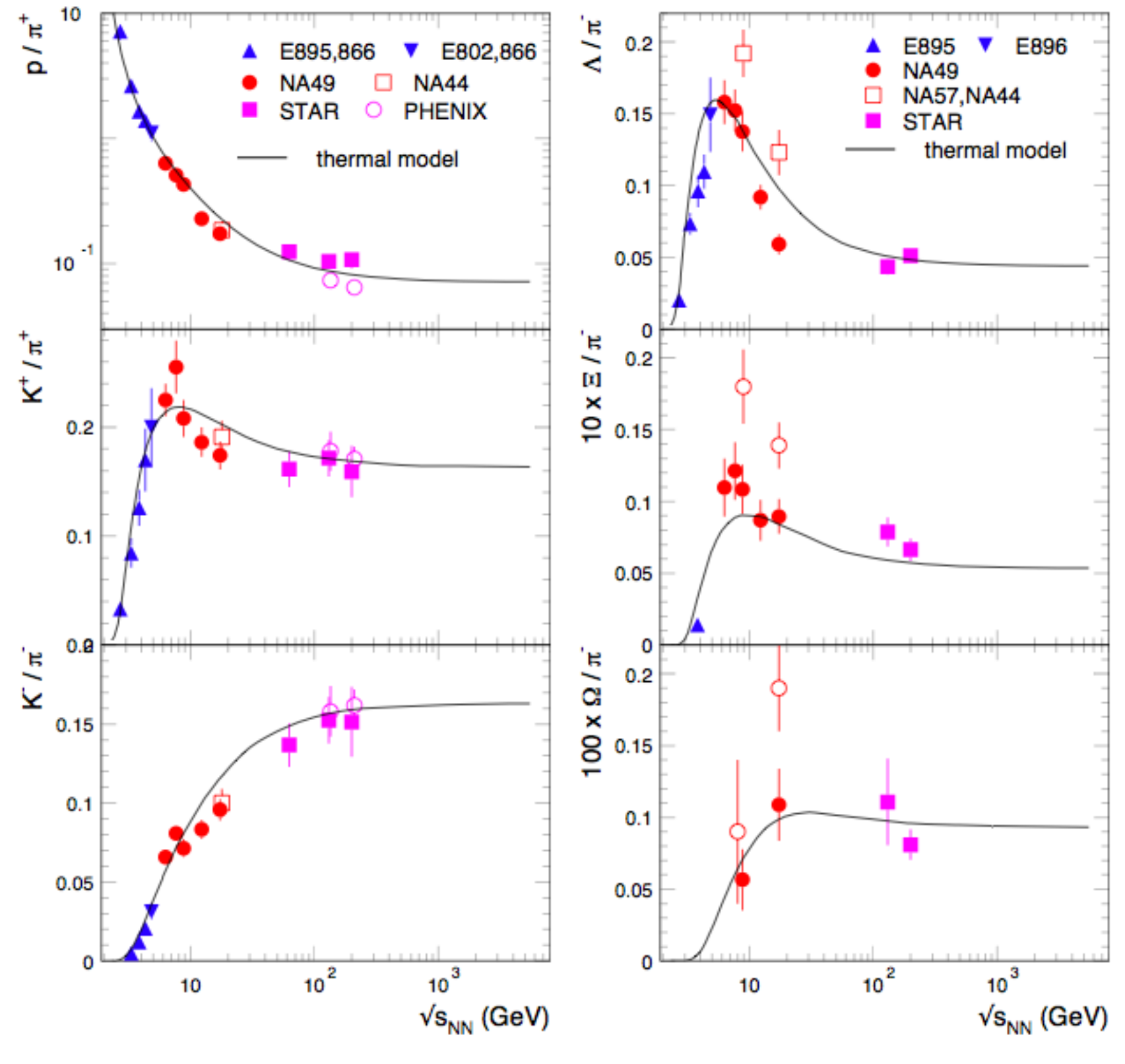} \vskip 0in }
 \caption{\label{fig_scales} Ratios of particle abudances.  }
 \label{marek}
 \end{figure}

The deconfinement transition should occur at some temperature fixed that is independent of $\mu_B$.
The Quarkyonic transition should occur roughly where $(\mu_B -M_N)/T \sim cons.$.   In Fig. \ref{freezeout}
the parameter for freezeout as determined by fits to particle abundances is compared to such simple model considerations.\cite{Andronic:2009gj}
Ratios of particle abundance should show singular behaviour as one approaches the triple point.  This indeed occurs, as shown in Fig. \ref{marek},\cite{Andronic:2005yp}
 The typical center of mass energy where this occurs is near $10~GeV$.  This suggest heavy ion experiments at  cernter of mass energies of this order and smaller may be able to access the physics of Happy Island.  
 
 \section{Summary}
 
 The physics of Happy Island is rich and unexplored.  Little is really known from theory, as speculative conclusions rely heavily on model considerations.  Reliable lattice gauge computation is difficult if not impossible to do at high density and low temperature.  On the experimental side we have hints of phase boundaries and a triple point not accounted for simply by the existence of a confinement-deconfinement phase transition.  What precisely to look for  is not known, and a phenomenology of Happy Island is yet non-existent.  Nevertheless the arguments for new physics in this region of the $\mu_B$-$T$ plane are compelling.  The possible existence of inhomogeneous phases might ultimately have interesting consequences for the physics of neutron stars.
 
 To infer the existence of Happy Island and Quarkyonic Matter, one has used a large number of colors approximation.  For mesons, it is widely believed that such an approximation is reasonable.  For baryons there are serious doubts.  The issue for us, whether or not quark loops affect the confinement potential is crucial  lattice Monte Carlo data and charmonium potential models suggest that in vacuum quark pairs provide little modification to a linear potential out to a distance of the order of a Fermi.  This suggests that at least for some range of baryon density, quark pairs provide little modification of the confining potential on distance scales corresponding to confinement.
 
 \section{Acknowledgements}

One of the pleasures of doing physics is the combination of imaginative thought and tangible realization embodied in both the ideas we study and the people who study them.   This meeting is in honor of  a a wonderful
friend and colleague, Walter Greiner.  Walter did much of the seminal work on the superheavy nuclei and the Island of Stability,
concepts that must have sounded as wild and speculative as the idea of a Happy Island for high density strongly interacting matter.  In this spirit of the imagination, my wife Alice and I have written the following poem for Walter: 

\begin{center}

{\bf An Ode to Dragons\\}
{\it For Walter\\}

\vspace{0.1 in}
{A dragon is a fearsome beast\\
With massive jaws and fiery breath;\\
And those who cannot match his strength\\
May hide themselves and wish his death\\

\vspace{0.1in}

A dragon is adventuresome\\
Unfazed by ergodicity.\\
He does not keep to precincts ruled\\
By order and simplicity.\\

\vspace{0.1in}

We see few dragons in these days\\
Yet might remember as a dream\\
A glimpse of huge wings in the sky\\
And scales with a metallic gleam...\\

\vspace{0.1in}

A dragon lives for endless years,\\
And many are the tales that tell\\
Of maidens rescued just in time--\\
Yet did they wish to end the spell?\\

\vspace{0.1in}

A dragon leads a lonely life\\
And could be kinder than most know,\\
For it is hard to be a friend\\
To one who see you as a foe.\\

\vspace{0.1in}

A child with no fear in his heart\\
May feel a lifting toward the sky\\
As far above, a dragon shares\\
Just how it feels to soar and fly.\\

\vspace{0.1in}

{\it by Alice and Larry McLerran\\2010}
}
\end{center}
 
\pagebreak

The research of  L. McLerran is supported under DOE Contract No. DE-AC02-98CH10886.
I also thank Debades Bandyopadhyay for the wonderful job he did in organizing this meeting.

\bibliographystyle{ws-procs9x6}
\bibliography{ws-pro-sample}

\end{document}